\documentclass[showpacs,amsmath,amssymb,aps,pra,10pt,reprint,superscriptaddress,showkeys]{revtex4-1}
\usepackage{graphicx}
\usepackage{bm}
\usepackage[breaklinks=true,colorlinks=true,linkcolor=blue,urlcolor=blue,citecolor=blue]{hyperref}
\usepackage{graphics}
\usepackage{dcolumn}
\usepackage{amsmath,amssymb}
\usepackage{mathdots}
\usepackage{natbib}
\usepackage{soul,color}
\usepackage{epstopdf}
\usepackage[note-name]{notes2bib}

\begin{document}

\title{Upper frequency limits for vortex guiding and ratchet effects}

\author{O. V.~Dobrovolskiy}
    \email{oleksandr.dobrovolskiy@univie.ac.at}
    \affiliation{Faculty of Physics, University of Vienna, 1090 Vienna, Austria}
    \affiliation{Physics Department, V. Karazin National University, 61077 Kharkiv, Ukraine}
\author{E. Begun}
    \affiliation{Physikalisches Institut, Goethe University, 60438 Frankfurt am Main, Germany}
\author{V. M. Bevz}
    \affiliation{Physics Department, V. Karazin National University, 61077 Kharkiv, Ukraine}
\author{R. Sachser}
    \affiliation{Physikalisches Institut, Goethe University, 60438 Frankfurt am Main, Germany}
\author{M. Huth}
    \affiliation{Physikalisches Institut, Goethe University, 60438 Frankfurt am Main, Germany}
\date{\today}

\begin{abstract}
Guided and rectified motion of magnetic flux quanta are important effects governing the magneto-resistive response of nanostructured superconductors. While at low ac frequencies these effects are rather well understood, their manifestation at higher ac frequencies remains poorly investigated. Here, we explore the upper frequency limits for guided and rectified net motion of superconducting vortices in epitaxial Nb films decorated with ferromagnetic nanostripes. By combining broadband electrical spectroscopy with resistance measurements we reveal that the rectified voltage vanishes at a \emph{geometrically defined frequency} of about 700\,MHz. By contrast, vortex guiding-related low-ac-loss response persists up to about $2$\,GHz. This value corresponds to the \emph{depinning frequency} $f_\mathrm{d}^\mathrm{s}$ associated with the washboard pinning potential induced by the nanostripes and exhibiting peaks for the commensurate vortex lattice configurations. Applying a sum of dc and microwave ac currents at an angle $\alpha$ with respect to the nanostripes, the angle dependence of $f_\mathrm{d}^\mathrm{s}(\alpha)$ has been found to correlate with the angle dependence of the depinning current. In all, our findings suggest that superconductors with higher $f_\mathrm{d}^\mathrm{s}$ should be favored for an efficient vortex manipulation in the GHz ac frequency range.
\end{abstract}


\maketitle

\section{Introduction}
Superconducting elements are important building blocks in quantum information processing \cite{Dev13sci}, circuit quantum electrodynamics \cite{Wal04nat}, and photon detectors for astrophysical applications \cite{Day03nat,Ale19prd}. They form the basis for superconducting qubits \cite{Cla08nat}, resonators \cite{Gol12sst}, and various Josephson \cite{Bar00boo} and Abrikosov fluxonic devices, such as filters \cite{Dob15apl}, rectifiers \cite{Vil03sci}, generators \cite{Dob18apl}, triodes \cite{Vla16nsr}, and bolometers \cite{Loe19acs}. All these devices are known to suffer from losses due to the motion of magnetic flux quanta. The vortices, unless pinned, increase noise and bit error rate in superconducting quantum interference devices (SQUIDs) \cite{Koe99rmp}, raise dark count rates in photon detectors \cite{Eng12prb} and reduce quality factors \cite{Son09apl} and power handling capabilities \cite{Che14apl}. Controlling the fluxon dynamics is therefore crucial for the performance of superconducting devices \cite{Lee99nat,Dob18nac}.

In this regard, guided and rectified motion of vortices are important effects allowing one to affect the magneto-resistive response of superconductors with nanoengineered vortex pinning sites \cite{Nie69jap,Lee99nat,Pas99prl,Dan00prb,Shk06prb,Dob10sst,Lee99nat,Ols01prl,Vil03sci,Von05prl,Luq07prb,Vel08mmm,Jin10prb,Shk14pcm,Wor12prb,Cer13njp,Sol14prb,Dob15met,Dob18nac,Dob17sst}. While vortex guiding implies a non-collinearity of the vortex velocity with the driving force exerted on them by the transport current \cite{Nie69jap,Pas99prl,Dan00prb,Shk06prb,Dob10sst}, rectified net motion of magnetic flux quanta occurs in systems lacking reflection symmetry---so-called vortex ratchets \cite{Lee99nat,Vil03sci,Vel08mmm,Jin10prb,Shk14pcm,Wor12prb,Cer13njp,Sol14prb,Dob15met}. At present, these effects in the dc- and low-frequency ac-driven regimes are rather well understood, see e.g. \cite{Plo09tas,Sil10inb,Dob17pcs} for reviews. In particular, vortex guiding and ratchet effects were investigated both experimentally \cite{Von05prl,Pry06apl,Dob17sst,Jin10prb,Cer13njp,Wor12prb,Dob15met} and theoretically \cite{Ols01prl,Shk06prb,Luq07prb,Shk14pcm}. At kHz ac frequencies the vortex dynamics is known to be frequency-independent \cite{Lar09pcm} and it begins to depend on the ac frequency in the MHz range \cite{Pry06apl,Jin10prb,Cer13njp,Sol14prb}. A weakening of the ratchet effect in nanopatterned superconducting films at several GHz was observed experimentally \cite{Jin10prb,Wor12prb,Dob15met}, and addressed theoretically \cite{Shk14pcm}. It was also demonstrated that rectified vortex motion can occur due to the asymmetry of the edge barriers of plain \cite{Pry06apl,Ali09njp} and nanopatterned \cite{Cer13njp} films. Finally, periodically arranged pinning sites were demonstrated to modify the vortex dynamics at high vortex velocities \cite{Sil12njp,Shk17prb,Dob19pra}, and a high-frequency ac stimulus was revealed to lead to the inhibition of vortex avalanches \cite{Awa11prb,Lar17pra} and stimulation of superconductivity in the vortex state \cite{Lar15nsr,Dob19rrl}.

Recently, the microwave response of systems composed of superconducting and ferromagnetic nanostructures has attracted especial attention \cite{Bel08prl,Gol18afm,Jeo18pra,Jeo18nam,Kim18prl,Rog19prm,Gol19pra,Dob19nph}. Due to the antagonistic spin ordering \cite{Lin15nph} the interplay of superconductivity and ferromagnetism leads to competing ground states, affecting the spin transport \cite{Kim18prl} and the dynamics of magnetic moment excitations \cite{Bel08prl,Gol18afm,Jeo18pra,Jeo18nam,Kim18prl,Rog19prm,Gol19pra,Dob19nph}. Exemplary phenomena emerging at microwave frequencies include the generation of superconducting pure spin currents \cite{Jeo18nam}, shortening of the quasiparticle charge-imbalance relaxation length across the superconducting transition temperature \cite{Jeo18pra}, modifications of the dispersion properties of spin waves \cite{Gol18afm,Gol19pra}, the formation of magnonic band structures \cite{Dob19nph}, as well as metamaterial properties \cite{Pim05prl}. In addition, trajectories of moving flux quanta can be imprinted in a soft magnetic layer \cite{Sha19met}, and guiding and ratchet effects are also being studied for whirl-like spin textures known as skyrmions \cite{Lew16inb,Max17prb,Che19prb}.

\clearpage
Here, by combining broadband electrical spectroscopy with dc voltage measurements, we show that the vortex ratchet effect in Nb films decorated with ferromagnetic Co nanostripes vanishes at a \emph{geometrically defined frequency} $f_\mathrm{r}$ of about $700$\,MHz. This frequency is related to the time needed for a vortex to travel between two neighboring Co nanostripes. By contrast, vortex-guiding-affected microwave excess losses persist up to about $2$\,GHz which corresponds to the vortex \emph{depinning frequency} $f_\mathrm{d}^\mathrm{s}$ \cite{Git66prl,Gol94prb,Jan06prb,Pom08prb,Zai03prb,Lar17pra} in the washboard pinning potential landscape induced by the Co nanostripe array. In addition, we observe maxima in the depinning frequency when the vortex lattice is commensurate with the periodic pinning potential landscape. By comparing the vortex guiding and ratchet effects for different current tilt angles with respect to the nanostripes, we deduce the angle dependence of the depinning frequency and find it to correlate with the angle dependence of the depinning current. Our results suggest that superconductors with stronger pinning (resulting in higher $f_\mathrm{d}^\mathrm{s}$) and denser pinning arrays (resulting in higher $f_\mathrm{r}$) should be favored for an efficient vortex manipulation in the GHz ac frequency range.

\section{Experimental}

\begin{figure}[tbh!]
    \centering
    \includegraphics[width=0.82\linewidth]{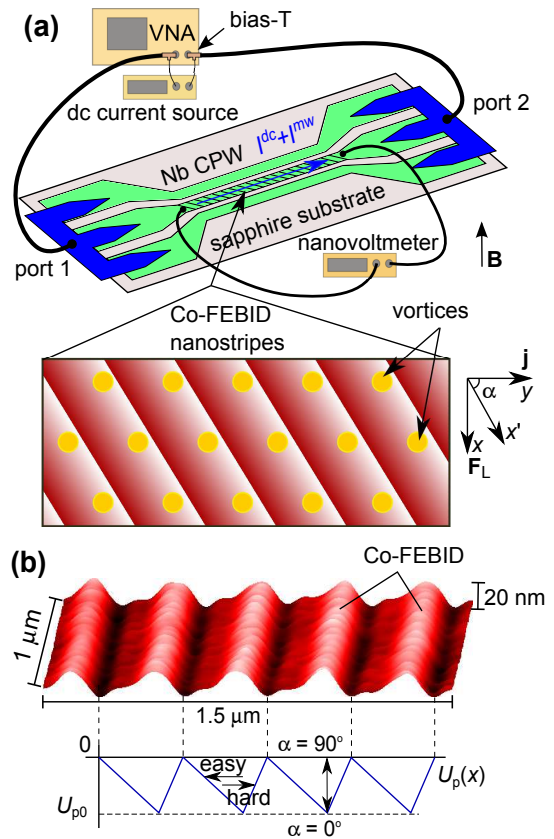}
    \caption{(a) Experimental geometry: dc and microwave ac currents $I^{dc}$ and $I^{mw}$ are applied to a Nb CPW decorated with an array of Co-FEBID nanostripes. The nanostripes are tilted at an angle $\alpha$ with respect to the CPW axis. The currents exert a Lorentz force $\mathbf{F}_L$ on vortices (denoted by yellow discs) acting along the $x$-axis. The vortex lattice is commensurate with the pinning landscape at 20\,mT. For oblique angles $\alpha\neq0^\circ, 90^\circ$, at sufficiently strong driving forces, vortices move in the guiding direction $x^\prime$ along the nanostripes. (b) AFM image of the Nb CPW decorated with Co-FEBID nanostripes (top) and inducing a pinning potential of the washboard type $U(x)$ with the depth $U_{p0}$ and the period $300$\,nm (bottom). With increase of $\alpha$, $U_{p0}$ decreases and vanishes at $90^\circ$. The vortex motion against the nanostripes' steep slopes is in the ``hard direction'' and it is in the ``easy direction'' against their gentle slopes.}
    \label{f1}
\end{figure}

The geometry of the experiment is shown in Fig. \ref{f1}(a). The samples are four nanopatterned Nb coplanar waveguides (CPWs) fabricated from a 56\,nm-thick epitaxial (110) Nb film sputtered by dc magnetron sputtering onto an a-cut sapphire substrate. In the sputtering process the film growth rate was about $1$\,nm/s, the substrate temperature was $T = 850^\circ$C, and the Ar pressure was $5\times10^{-3}$\,mbar \cite{Dob12tsf}. The CPWs were fabricated by photolithography in conjunction with Ar etching. The samples are characterized by the superconducting transition temperature $T_\mathrm{c} = 8.05$\,K, the upper critical field at zero temperature $H_{c2}(0) \simeq 1.4$\,T, and the coherence length $\xi(0)\approx 15$\,nm. The $H_{c2}(0)$ value was deduced from fitting the dependence $H_{c2}(T)$ to the phenomenological law $H_{c2}(T) = H_{c2}(0) [1-(T/T_c)^2]$. The standard relation $\xi(0) = [\Phi_0/(2\pi H_\mathrm{c2}(0))]^{1/2}$ was used to estimate the superconducting coherence length. The central conductor of the CPWs is $100\,\mu$m long and $30\,\mu$m wide while the gap between the center and ground conductors is $12\,\mu$m, to match the 50\,$\Omega$ impedance of the feedline. The voltage contacts were wire-bonded on top of the center conductor of each CPW.

The central conductor of each CPW was decorated with parallel, $300$\,nm-spaced Co-based nanostripes directly written by focused electron beam induced deposition (FEBID) \cite{Hut18mee,Dob15bjn,Dob19pra}. We refer to Appendix for further details on the FEBID process. The Co nanostripe array induces a pinning potential of the washboard type in the Nb film via local suppression of the superconducting order parameter via the proximity effect. Inspection by atomic force microscopy (AFM), see Fig. \ref{f1}(b), confirms the targeted $300$\,nm-periodicity of the array of nanostripes and the desired asymmetry of their cross-section necessary to break the pinning strength symmetry under current polarity reversal. The Co nanostripes have a height of 20\,nm and a half-height width $b$ of about 100\,nm. To investigate the different regimes of guided and rectified motion of vortices, the nanostripes were tilted at angles $\alpha=0^\circ$, $30^\circ$, $60^\circ$, and $90^\circ$ with respect to the current direction. Specifically, the sum of dc and ac currents applied along the $y$-axis in a magnetic field $\mathbf{B} \equiv \mathbf{B}_z= \mu_0 \mathbf{H}$ exerts on a vortex a Lorentz force per unit length $\mathbf{F}_L = \Phi_0 [\mathbf{j}\times \mathbf{z}]$ acting along the $x$-axis. Here, $\mathbf{j}$ is the electric current density and $\mathbf{z}$ is the unit vector in the $z$ direction.

\clearpage
Combined broadband microwave and dc voltage measurements were done at $0.8T_\mathrm{c} = 6.44$\,K with magnetic field $\mathbf{H}$ directed perpendicular to the film surface. The dc voltage $V$ and the absolute value of the forward transmission coefficient $S_{21}$ were measured as a function of the dc bias current value $I^{dc}$. To simplify the notation, in the following sections we will omit the superscript ``dc'' in $I^{dc}$. The microwave signal was generated and analyzed by an Agilent E5071C vector network analyzer (VNA). To present the vortex-related microwave loss, in what follows we use the notation of insertion loss, $IL$, deduced by the standard relation $IL \equiv  - S_{21}(f,P,H)/S_{21}(f,P,H_{\mathrm{ref}})$ \cite{Jin10prb,Wor12prb,Che14apl,Lar15nsr,Sol14prb}, where $S_{21}(f,P,H_\mathrm{{ref}})$ is the frequency- and microwave-power-dependent reference loss at $H_\mathrm{ref} >H_{c2}$.

\section{Results}

\begin{figure}[t!]
    \centering
    \includegraphics[width=1\linewidth]{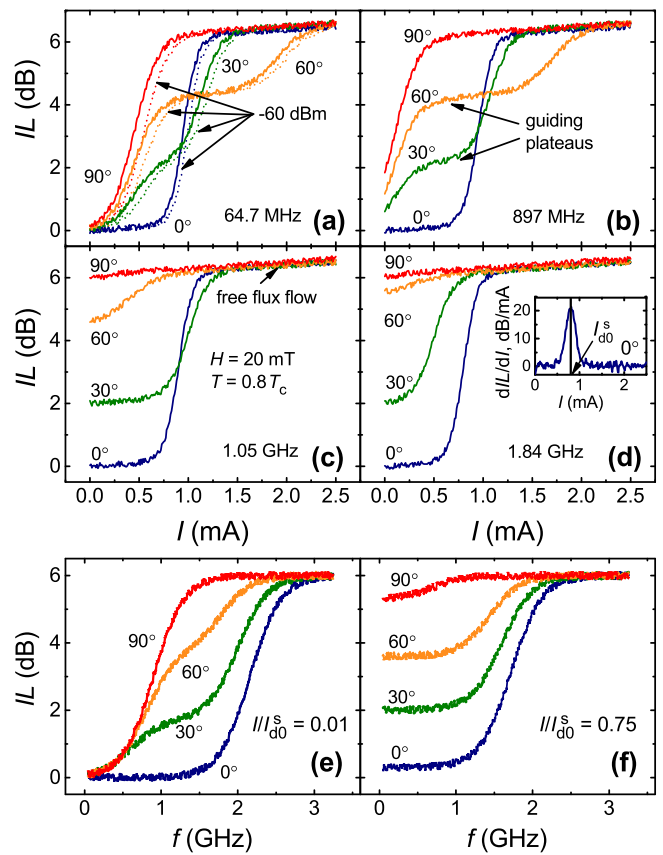}
    \caption{(a)-(d) Vortex-related insertion loss $IL$ as a function of the dc current value in the presence of an ac current at a power level of $-20$\,dBm (solid lines) for a series of frequencies, as indicated.
    In (a), the dashed lines correspond to the same measurement of $IL(I)$ at a power level of $-60$\,dBm.
    (e) and (f) $IL(f)$ for two dc current values normalized to the depinning current $I^\mathrm{s}_\mathrm{d0}$ whose definition is shown in the inset of (d)
    depicting the current dependence of the derivative of $IL$ with respect to current for $\alpha = 0^\circ$.
    In all panels $T = 0.8T_\mathrm{c}$ and $H = 20$\,mT.
    }
    \label{f2}
\end{figure}

Figure \ref{f2} displays the vortex-related microwave loss with increase of the dc current of positive polarity at $H = 20$\,mT and four ac frequencies. At this field value the vortex lattice is commensurate with the nanostripe array, as illustrated in Fig. \ref{f1}(a). For other angles, to minimize the energy, the vortex lattice is rotated to adopt itself to the ``channels'' of the periodic pinning potential \cite{Gui14nph}. The data in Fig. \ref{f2}(a) were acquired at $64.7$\,MHz, as is exemplary for the quasistatic regime, at an ac power of $- 60$\,dBm ($1$\,nW, dashed lines) and $- 20$\,dBm ($10\,\mu$W, solid lines). The panels (b) to (d) in Fig. \ref{f2} illustrate the increase of the insertion loss with increase of the frequency up to $1.84$\,GHz, while panels (e) and (f) present the insertion loss frequency dependences at very small and moderately large dc current values.

\begin{figure}
    \centering
    \includegraphics[width=0.72\linewidth]{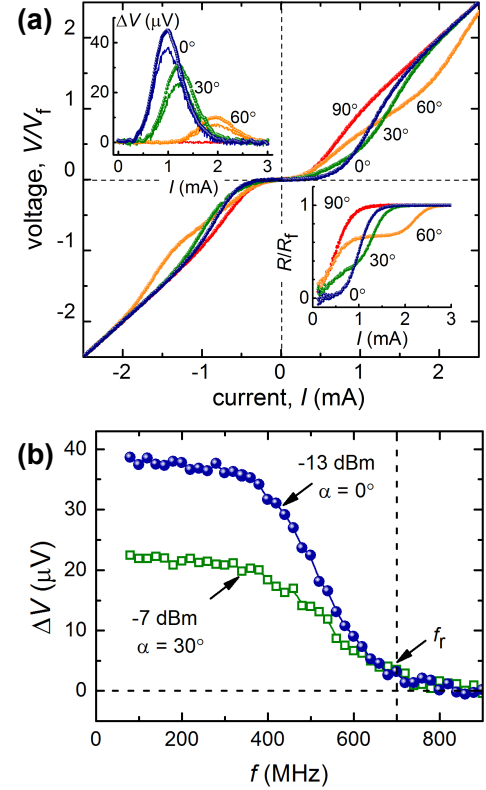}
    \caption{(a) $I$-$V$ curves of all samples in the absence of an ac stimulus. Lower inset: Current dependence of the normalized resistance. Upper inset: Symbols: Difference voltage $\Delta V = V(I) + V(-I)$ plotted by adding the right-hand and left-hand branches of the respective $I$-$V$ curves in the main panel. Solid lines: Rectified dc voltage appearing in the presence of an ac current with $f=64.7$\,MHz. (b) Frequency dependence of the rectified voltage $\Delta V(f)$ for the ac power levels and angles $\alpha$, as indicated. The rectified voltage vanishes above $f_\mathrm{r}\approx 700$\,MHz. In all panels $T = 0.8T_\mathrm{c}$ and $H = 20$\,mT.}
    \label{fCVC}
\end{figure}

We first consider the reference dashed curves in Fig. \ref{f2}(a) acquired in the weak-ac-drive limit ($-60$\,dBm). The curves have a smeared step functional shape for $0^\circ$ and $90^\circ$, while those for $30^\circ$ and $60^\circ$ have a double-step one. The presence of two smeared steps in the curves at $- 60$\,dBm in Fig. \ref{f2}(a) can be understood as a consequence of the coexistence of the intrinsic pinning in the Nb film and the anisotropic pinning induced by the Co nanostripes. Namely, as the dc current increases, vortices first have to overcome the barriers of the intrinsic pinning characterized by the depinning current $I_\mathrm{d}^\mathrm{i}$. This intrinsic pinning is weak and isotropic, as concluded from $I_\mathrm{d}^\mathrm{i}(30^\circ) \approx I_\mathrm{d}^\mathrm{i}(60^\circ) \approx I_\mathrm{d}^\mathrm{i}(90^\circ)= 0.5$\,mA. Here, $I_\mathrm{d}^\mathrm{i}$ is determined at the maximum of the derivative $d[IL(I)]/dI$ and we use the same definition for $I_\mathrm{d}^\mathrm{s}$ associated with the washboard pinning potential induced by the Co nanostripes, see the inset of Fig. \ref{f2}(d). The motion of vortices at $0^\circ$ is most efficiently impeded by the nanostripes and this is why the low-dissipative regime is preserved up to $I_\mathrm{d0}^\mathrm{s} \approx 0.97$\,mA. Importantly, $I_\mathrm{d0}^\mathrm{s}\equiv I_\mathrm{d}^\mathrm{s}(0^\circ) < I_\mathrm{d}^\mathrm{s}(30^\circ) < I_\mathrm{d}^\mathrm{s}(60^\circ) \approx 2$\,mA, which is a fingerprint of the vortex guiding effect extensively investigated in the dc-driven regime \cite{Nie69jap,Sil10inb,Dob10sst,Dob11pcs} and illustrated further in Fig. \ref{fCVC}(a). This effect consists in that vortices move easier along the pinning channels induced by the Co stripes than overcome the associated pinning potential barriers. Given that only the Lorentz force component acting perpendicularly to the Co stripes drives the vortices across them, one has $I_\mathrm{d}^\mathrm{s}(\alpha) = I_\mathrm{d}^\mathrm{s}(0^\circ)/\cos\alpha$ \cite{Shk06prb}. Accordingly, with $I_\mathrm{d}^\mathrm{s}(0^\circ)\approx 1$\,mA deduced from Fig. \ref{fCVC}(a) one obtains $I_\mathrm{d}^\mathrm{s}(30^\circ)= I_\mathrm{d}^\mathrm{s}(0^\circ)/\cos30^\circ \approx 1.16$\,mA and $I_\mathrm{d}^\mathrm{s}(60^\circ) = I_\mathrm{d}^\mathrm{s}(0^\circ)/\cos60^\circ \approx 2$\,mA that agrees with the experimental observation. At the same time, at $\alpha =90^\circ$ vortices move along the nanostripes and do not experience the washboard pinning potential which is why $I_\mathrm{d}^\mathrm{s}(90^\circ)$ can not be defined. The vortex-related insertion loss for $30^\circ$ and $60^\circ$ exhibits plateaus at $0.7\,\mathrm{mA} \lesssim I\lesssim 0.9\,\mathrm{mA}$ and $0.7\,\mathrm{mA} \lesssim I\lesssim 1.5\,\mathrm{mA}$, respectively. In this regime, $\mathbf{v} \nparallel \mathbf{F}_L$ and vortices move along the nanostripes \cite{Sil10inb,Shk06prb}. Finally, at larger dc currents the vortices overcome barriers of the pinning potential induced by the nanostripes, and the regime of free flux flow sets in with $\mathbf{v} \parallel \mathbf{F}_L$. In the presence of the $-20$\,dBm microwave excitation in Fig. \ref{f2}(a-d), all crossovers shift towards smaller dc currents.

The evolution of the vortex-related insertion loss with increase of the ac frequency is illustrated in Fig. \ref{f2}(b)-(d). Specifically, between 64.7\,MHz and 897\,MHz the addition of the ac current shifts the crossovers related to $I_\mathrm{d}^\mathrm{i}$ towards smaller currents whereas the crossovers related to $I_\mathrm{d}^\mathrm{s}$ remain almost unaffected, see Fig. \ref{f2}(b). With increase of the ac frequency to $1.05$\,GHz, as shown in Fig. \ref{f2}(c), the curve for $\alpha = 60^\circ$ is shifted to the left and the order of the crossovers at $I_\mathrm{d}^\mathrm{s}$ for different angles becomes interchanged. With further increase of $f$ to $1.84$\,GHz the most notable change happens to the curve with $\alpha = 30^\circ$, for which now $I_\mathrm{d}^\mathrm{s}(30^\circ) < I_\mathrm{d}^\mathrm{s}(0^\circ)$. Finally, when $f$ approaches 2.22\,GHz (not shown), $IL$ reaches about $6$\,dB at all $\alpha$ and is very weakly increasing with further increase of the dc current value.

At $I/I^\mathrm{s}_\mathrm{d0}=0.01$, which is exemplary for the regime of weak dc currents, the frequency dependence of the vortex-related insertion loss is depicted in Fig. \ref{f2}(e). The curves for $\alpha = 0^\circ$ and $90^\circ$ qualitatively resemble the well-known results of Gittleman and Rosenblum (GR) \cite{Git66prl,Pom08prb}. In the GR model, the frequency dependence of the microwave power absorbed by vortices at fixed magnetic field and temperature exhibits a single-step crossover from the weak dissipation at low frequencies, where the pinning forces dominate, to the strong dissipation at high frequencies, where the frictional forces prevail. Quite distinct from these GR-like curves, \emph{double-step} crossovers at $\alpha = 30^\circ$ and $60^\circ$ are clearly seen and allow one to introduce a characteristic depinning frequency for each crossover, at the maximum of its frequency derivative. By contrast, at close-to-depinning currents, which is exemplified in Fig. \ref{f2}(f) for $I/I^\mathrm{s}_\mathrm{d0}=0.75$, single-step crossovers are observed for all $\alpha$ and the difference in the insertion loss at low and high frequencies becomes smaller with increasing $\alpha$. This can be understood as a consequence of the effective weakening of the effective periodic pinning potential with increase of $\alpha$ as the driving force in this case is counterbalanced by the pinning force component which is proportional to $\cos\alpha$.

Finally, we present the current-voltage ($I$-$V$) curves for all samples in Fig. \ref{fCVC}(a) where one can see the crossovers from the pinned and guided regimes to free flux flow in the absence of an ac current. With the superscripts ($+$) and ($-$) denoting the respective dc current polarity, we note that $I^{\mathrm{s}+}_\mathrm{d} > |I^{\mathrm{s}-}_\mathrm{d}|$ for all $\alpha \neq 90^\circ$ due to the asymmetry of the nanopattern. For $\alpha = 90^\circ$ the depinning current $I^{\mathrm{i}+}_\mathrm{d} = |I^{\mathrm{i}-}_\mathrm{d}|$, as expected, as the effective asymmetry of the pinning landscape vanishes in this case. The lower inset of Fig. \ref{fCVC}(a) displays the current dependences of the normalized resistance $R(I)/R_\mathrm{f}$ which mimics the shape of the curves in Fig. \ref{f2}(a). Here, $R_\mathrm{f}$ is the flux-flow resistance.

In the absence of dc current, application of an ac current with $f=64.7$\,MHz results in the appearance of a rectified voltage $\Delta V$ of the order of $10\,\mu$V for all $\alpha\neq 90^\circ$, see the upper inset in Fig. \ref{fCVC}(a). This exception is expected as at $\alpha =90^\circ$ the vortices move along the nanostripes and the asymmetry of the nanostripe slopes does not affect vortex motion. The rectified voltage $\Delta V(I)$ has a dome-like shape as a function of the ac current amplitude and its magnitude decreases with increase of $\alpha$. For comparison, the reference curves obtained by plotting the difference voltage between the right-hand $V(I^+)$ and left-hand $V(I^-)$ branches of the dc-driven $I$-$V$ curve in the main panel in Fig. \ref{fCVC}(a) are shown by solid lines and resemble the behavior of the $\Delta V(I)$ curves obtained in response to the $64.7$\,MHz-frequency ac drive. Importantly, the magnitude of the rectified voltage decreases with increase of the ac frequency and eventually vanishes above $f_\mathrm{r}\approx 700$\,MHz, as can be concluded from Fig. \ref{fCVC}(b).

We note that $\Delta V (I)$ attains its maximum at $I_\mathrm{d}^\mathrm{s}(0^\circ) \approx 1$\,mA and $I_\mathrm{d}^\mathrm{s}(30^\circ) \approx 1.2$\,mA, see the upper inset in Fig. \ref{fCVC}(a). Accordingly, the evolution of the maximal $\Delta V(f)$ is shown in Fig. \ref{fCVC}(b) for the ac currents $1$\,mA and $1.2$\,mA (corresponding to $-13$\,dBm and $-11.5$\,dBm) for $\alpha = 0^\circ$ and $30^\circ$, respectively. Because of the asymmetry of the $I$-$V$ curves in Fig. \ref{fCVC}(a), we introduce the averaged voltage $\overline{V}= [V(I^+) + |V(I^-)|]/2$ such that $1$\,mA corresponds to $\overline{V} \approx 0.81$\,mV for $\alpha=0^\circ$ and $1.2$\,mA corresponds to $\overline{V} \approx 0.99$\,mV for $\alpha=30^\circ$. Employing the standard relation for the vortex velocity $v = \overline{V}/BL$, where $B=20$\,mT and $L=100\,\mu$m is the distance between the voltage contacts, we deduce $v(0^\circ, 1\,\mathrm{mA}) \approx 405$\,m/s and $v(30^\circ, 1.2\,\mathrm{mA}) \approx 495$\,m/s. Following Refs. \cite{Wor12prb,Lar17pra}, we estimate the displacement of vortices during one ac period as $2d(0^\circ) = v/f_\mathrm{r}$ and $2d(30^\circ) = v\cos30^\circ/f_\mathrm{r}$, yielding $d(0^\circ) \approx 290$\,nm and $d(30^\circ) \approx 310$\,nm, respectively. This suggests that the rectified voltage vanishes as soon as the amplitude of vortex oscillations $d$ becomes smaller than the nanopattern period $a$.

\section{Discussion}
We proceed to a discussion of the experimental findings, namely (i) the two-step increase of the insertion loss $IL$ at $\alpha\neq0^\circ,90^\circ$, and (ii) the different upper frequencies for the existence of the rectified voltage (about $700$\,MHz) and the low-dissipative microwave response (about $2$\,GHz at $\alpha = 0^\circ$).

First, it should be recalled that an increase of the vortex-state microwave loss in superconductors is known to occur at the depinning frequency $f_\mathrm{d}$ \cite{Git66prl,Gol94prb,Jan06prb,Pom08prb,Zai03prb,Lar17pra}. This frequency has the physical meaning of a crossover frequency from the low-frequency regime, where the pinning forces dominate and the vortex response is weakly dissipative, to the high-frequency regime, where the frictional forces prevail and the response is strongly dissipative. The experimental data in Fig. \ref{f2} point to the presence of \emph{two different} depinning frequencies in our system, $f_\mathrm{d}^\mathrm{i}$ and $f_\mathrm{d}^\mathrm{s}$. The frequency $f_\mathrm{d}^\mathrm{i}$ can be attributed to the intrinsic pinning in the Nb film while $f_\mathrm{d}^\mathrm{s}$ to the periodic pinning induced by the Co nanostripes.

\begin{figure}[t!]
    \centering
    \includegraphics[width=0.95\linewidth]{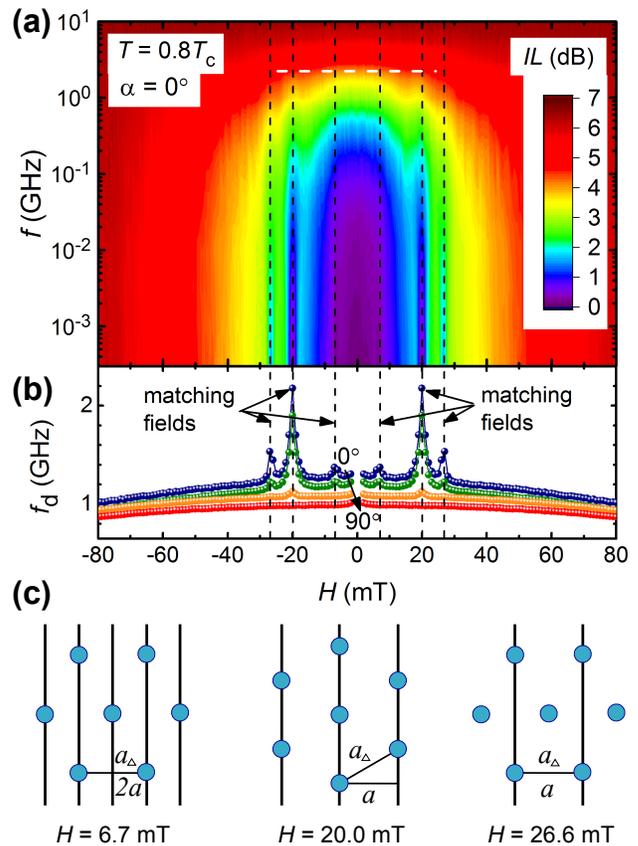}
    \caption{(a) Insertion loss as a function of the magnetic field and the ac frequency for $\alpha = 0^\circ$. The horizontal dashed line indicates the upper frequency $f \approx2$\,GHz above which the matching dips vanish.
    (b) Depinning frequency $f_\mathrm{d}$ as a function of the magnetic field at $T = 0.8T_\mathrm{c}$ in the absence of dc current for all angles $\alpha$.
    (c) Vortex lattice configurations at the three matching fields corresponding to the peaks in $f_\mathrm{d}(H)$ in panel (b).}
    \label{fSvH}
\end{figure}

To give further insights into the angle dependences of the depinning frequencies, the insertion loss has been measured in the absence of dc current in the weak-ac-drive regime ($-60$\,dBm) for all samples at magnetic fields varying from $-80$ to $80$\,mT. The data for $\alpha =0^\circ$ are shown in Fig. \ref{fSvH}(a). Namely, while the insertion loss increases with increase of the number of vortices (which is proportional to $H$), two minima are observed at $20$\,mT and $26.6$\,mT on the background of this increase. As mentioned before, $20$\,mT is a matching field value at which the vortex lattice is commensurate with the $300$\,nm-periodic nanostripe array. The arrangement of vortices at $20$\,mT in the pinning nanolandscape is shown in Fig. \ref{f1}(a) for the assumed triangular vortex lattice with lattice parameter $a_\bigtriangleup = (2\Phi_0/B\sqrt{3})^{1/2}$ and the matching condition $a_\bigtriangleup = 2a/\sqrt{3}$. In this configuration, all vortices are pinned by the nanostripes and there is no interstitial vortices. Since all vortices experience the same local pinning forces, the pinning efficiency at the matching field is maximal. We also note that no matching minima in the microwave loss are observed at frequencies above approximately $2$\,GHz, as will be discussed in what follows.

The dependences of the depinning frequency on the magnetic field in the absence of dc current for all samples are shown  in Fig. \ref{fSvH}(b). Namely, $f_\mathrm{d}(H)$ exhibits maxima (best seen at $0^\circ$) at $6.7$\,mT, $20$\,mT, and $26.6$\,mT and it reaches $2.16$\,GHz at $20$\,mT. Here, the  fields $6.7$\,mT and $26.6$\,mT are further matching fields with the matching condition $a_\triangle = na$ and the order number $n = 2$ and $n = 1$, respectively. We refer to Fig. \ref{fSvH}(c) for sketches of the vortex lattice configurations. The data exhibit a systematic decrease with increase of $\alpha$ and a weak decrease with increase of $H$. With increase of $\alpha$ the magnitude of the matching peaks decreases and $f_\mathrm{d}$ values converge to $1$\,GHz at large $\alpha$. The very weak dependence of the depinning frequency on $H$ can be explained by flattening of the previously observed dependence \cite{Jan06prb} following the law $f_\mathrm{d} = f_\mathrm{d}(T,0,I)[1 - (H/H_{c2})^2]$. This flattening of $f_\mathrm{d}(H)$ in our experiment is because of magnetic fields being much smaller than the upper critical field $H_{c2}(0.8T_\mathrm{c})  \approx0.5$\,T.

\begin{figure}[t!]
    \centering
    \includegraphics[width=0.68\linewidth]{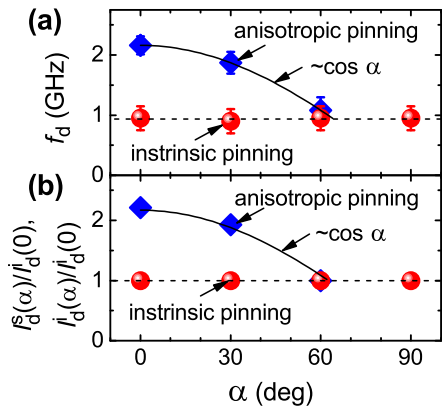}
    \caption{Angle dependences of the depinning frequencies and the normalized depinning currents for the intrinsic and anisotropic pinning at $T = 0.8T_c$ and $B = 20$\,mT.
    Solid lines are fits proportional to $\cos\alpha$.
     }
    \label{f6}
\end{figure}

The existence of two depinning frequencies in the decorated Nb films allows us to explain the evolution of the guiding and ratchet effects with increase of the ac frequency as follows. With increase of the current tilt angle $\alpha$ with respect to the nanostripes, (i) the \emph{depth} of the effective pinning potential \emph{is decreasing} while (ii) its \emph{width is increasing}. Here, we mean the effective width of the pinning potential in the direction of the driving force, since it is this effective width which is of primary importance for the excess ac loss associated with the oscillation of vortices at the bottoms of the pinning potential wells induced by the Co nanostripes. Accordingly, panels (a) and (b) in Fig. \ref{f6} illustrate a correlation of the depinning frequency with the depinning current for each of the pinning types in the investigated system. Thus, for the background isotropic pinning both, the depinning current and frequency are independent of $\alpha$. By contrast, both, the depinning current and frequency for the washboard pinning potential induced by the Co stripes decrease proportional to $\cos\alpha$ with increasing $\alpha$ which can be understood as a consequence of weakening of the normal component of the pinning force impeding the vortex motion.

From previous works it is known that the depinning frequency depends on temperature $f_\mathrm{d}\equiv f_\mathrm{d}(T,H,I) = f_\mathrm{d}(0,H,I)[1 - (T/T_{c})^4]$ \cite{Zai03prb} and magnetic field $f_\mathrm{d} = f_\mathrm{d}(T,0,I)[1 - (H/H_{c2})^2]$ \cite{Jan06prb}. Furthermore, it has been revealed to depend on the dc bias current as $f_\mathrm{d} = f_\mathrm{d}(T,H,0)[1- (I/I_\mathrm{d})^{3/2}]^{2/3}$ and $f_\mathrm{d} = f_\mathrm{d}(T,H,0)[1- (I/I_\mathrm{d})^{4}]^{1/4}$ for the ``easy'' and ``hard'' direction of the vortex motion, respectively \cite{Dob17nsr}, for washboard pinning landscapes with an asymmetry which is very close to the one used here. From $f_\mathrm{d} (0.8T_\mathrm{c}, 0.04H_{c2}) \approx 2.16$\,GHz deduced from Fig. \ref{fSvH} we estimate its zero-temperature value $f_\mathrm{d}(0, 0.04H_{c2}) \approx 3.6$\,GHz. This estimate correlates reasonably well with our previous experiments on Nb film with nanogroove arrays milled by focused ion beam \cite{Dob15apl}.

We emphasize that with an increase of the ac frequency, the rectified voltage $\Delta V$ vanishes at about $700$\,MHz, while the low-dissipative microwave response persists up to about $2\,$GHz. To explain the difference between these two frequencies, we compare our experiment with experiment \cite{Wor12prb} on YBCO films and outline their distinctive features. Namely, in Ref. \cite{Wor12prb} the upper frequency limits for vortex guiding and the ratchet effect were investigated on YBCO thin films patterned with antidot arrays. The triangular shape and arrangement of the antidots was designed to break the symmetry of the pinning potential landscape for vortex motion \emph{along} the antidot rows. In that work, the rectified ratchet voltage persisted up to essentially higher frequencies (about $8$\,GHz) as compared to the frequency at which the guiding-induced microwave loss vanished (about $2$\,GHz). Due to the discontinuity of the YBCO film with antidots, it was supposed \cite{Wor12prb} that flux transport at higher frequencies takes place via phase slips through the antidots isthmuses, each acting as a narrow superconducting channel. That suggestion explained the absence of a contribution to the microwave loss above $2$\,GHz while the rectified voltage persisted up to at least $8$\,GHz.

In contradistinction to that experiment, the order of the upper frequency limits is inverse for the the rectified voltage (about $700$\,MHz) and the crossover to the high-microwave-loss state (about $2$\,GHz) in our work. We note that in our experiment the film is continuous, as it is not split into narrow superconducting channels, and the symmetry of the pinning potential landscape is broken by nanopatterning in the direction \emph{transverse} to the guiding direction of the Co stripes. Accordingly, the role of the two length scales responsible for the formation of the rectified voltage and the vortex-related microwave loss are notably different from that in Ref. \cite{Wor12prb}. Namely, these length scales are the nanostructure period $a=300$\,nm for the rectified voltage and the full width at half height $b\approx100$\,nm as an estimate for the width of the pinning potential well determining the crossover from the low-dissipative to the highly-dissipative regime. Recalling that the vortex ratchet effect is a net effect implying that vortices visit \emph{more than one pinning site (nanostripe) during each ac cycle} we emphasize that guided vortex motion means that vortices move \emph{along one and the same pinning site (nanostripe)}. Accordingly, with an increase of the ac frequency, first the frequency is reached at which vortices have not enough time to get to the neighboring nanostripes and this corresponds to the vanishing of the rectified voltage. With further increase of the frequency, vortices have not enough time to reach the regions with strong pinning forces. These regions in our system are the film areas under the Co stripe slopes inducing the maximal gradients of the pinning potential. The about threefold difference between the two frequencies nicely corresponds to the ratio $a/b\approx 3$ of the employed nanostructure.

\clearpage
Finally, the general significance of the obtained results should be outlined.

(i) \emph{Ratchet effect and implications for vortex diodes}. While the investigated system can be used as a rectifier up to $700$\,MHz, the relation between the upper frequency of the ratchet voltage and the nanostructure period, $f_\mathrm{r} \sim v/a$, suggests that pinning arrays with a smaller period should result in higher $f_\mathrm{r}$. Since the current state of technology allows for the fabrication of sub-$100$\,nm-periodic arrays of pinning sites \cite{Aic19anm}, one can anticipate vortex diodes with cut-off frequencies above $10$\,GHz on the basis of superconducting films featuring small vortex core sizes in conjunction with high vortex velocities, such as e.g. superconducting cuprates \cite{Wor12prb}. At the same time, the observed correlation of the depinning frequency with the depinning current related to the anisotropic pinning potential, $f_\mathrm{d} \sim I_\mathrm{d}^\mathrm{s}$, means that the low-loss microwave response should persist up to higher frequencies for superconductors with strong periodic pinning.

(ii) \emph{Guiding effect and implications for microwave filters}. An important feature of the microwave response of guided magnetic flux quanta is a two-step crossover from the low-dissipative state to the highly dissipative state at oblique current tilt angles $\alpha \neq 0^\circ, 90^\circ$ with the respect to the guiding channels of the washboard pinning potential. In addition to the departure from the widely used GR model \cite{Git66prl} implying a single-step crossover for superconductors with one type of pinning, this feature not only allows for the development of microwave cut-off filters with tailored frequency roll-offs, but it can be used as a fundamental fingerprint of the presence of pinning of different strengths in the investigated system.

(iii) \emph{Matching peaks in the depinning frequency}. The coexistence of two pinning types of superconductors is reflected in the presence of two characteristic (depinning) frequencies in the microwave response. The observed strong dependence of the depinning frequency on the magnetic field near the matching values (about $250$\,MHz/mT for the fundamental matching field) can be suggested as a sensitive approach for the determination of commensurate vortex lattice configurations. This approach is expected to be especially valuable at lower temperatures for superconductors with strong pinning, in which dc resistance measurements at low currents result in voltages below the noise floor of the setup for resolving the matching peculiarities while at high dc currents the vortex motion causes a strong overheating of the superconductor.

(iv) \emph{Implications for hybrid devices exploiting vortex and spin-wave physics}. There is currently great interest in the microwave properties of hybrid superconductor-ferromagnet systems \cite{Gol18afm,Gol19pra,Jeo18pra,Jeo18nam}. While the Co stripes in our work are used solely as guiding channels for Abrikosov vortices, such an array is expected to behave as a magnonic crystal for spin waves with peculiar Bloch-like band structures in the GHz frequency range \cite{Chu17jpd}. While recently it was demonstrated that Co-based FEBID stripes can be used as magnonic conduits \cite{Dob19ami} and the interaction of Abrikosov vortices with spin waves leads to the formation of forbidden-frequency gaps in the magnon spectrum \cite{Dob19nph}, one may expect that the spin wave transmission in such a hybrid system will be modified by guided and rectified vortex motion. The tunability of $f_\mathrm{d}$ and $f_\mathrm{r}$ should allow then, in principle, for the discrimination of the characteristic features in the microwave response related to the superconducting and magnetic systems.

\section{Conclusion}
To summarize, we have studied by microwave spectroscopy and electrical voltage measurements the guided and rectified motion of magnetic flux quanta in Nb films decorated with Co nanostripes. Excess microwave loss due to vortices guided by the nanostripes is observed at ac frequencies up to about $2$\,GHz, while the rectified ratchet voltage vanishes already at about $700$\,MHz. In the investigated system, vortex guiding and the ratchet effect ensue on the background of the competition between the intrinsic weak pinning in the Nb films and the strong periodic pinning induced by the Co nanostripe array. Variation of the nanostripe tilt angle with respect to the current direction allows one to distinguish between two depinning frequencies associated with both pinning types in the samples. In particular, while $f_\mathrm{d}^\mathrm{i} \sim 1\,$GHz has been revealed to be smaller than $f_\mathrm{d}^\mathrm{s} \sim 1-2\,$GHz and independent of $\alpha$, an angle dependence of $f_\mathrm{d}^\mathrm{s} (\alpha) = f_\mathrm{d}^\mathrm{s}(0^\circ)\cos\alpha$ has been observed. This dependence correlates well with the angle dependence of the depinning current, $I_\mathrm{d}^\mathrm{s}(\alpha) = I_\mathrm{d}^\mathrm{s}(0^\circ)\cos\alpha$, associated with the crossover from the guided vortex motion along the nanostripes to the regime of free flux flow. In all, the obtained results suggest that superconductors with stronger pinning (resulting in higher $f_\mathrm{d}^\mathrm{s}$) and denser pinning arrays (resulting in higher $f_\mathrm{r}$) should be favored for an efficient vortex manipulation in the GHz ac frequency range.

\section*{Appendix}

\textbf{Fabrication of Co nanostripes}. The Nb CPWs were decorated by Co-based nanostripes by focused electron beam induced deposition (FEBID). FEBID was done in a high-resolution scanning electron microscope (SEM, FEI Nanolab 600) employing the precursor gas Co$_2$(CO)$_8$. This technique relies upon the dissociation of the metal-organic precursor in the focus of the electron beam into a permanent deposit and volatile components, in accordance with the pre-defined pattern. FEBID of Co was done with the 3\,kV/90\,pA beam parameters and 4200 beam passes. The pitch was $10$\,nm, the dwell time was $1\,\mu$s, and the process pressure was $1.1\times10^{-5}$\,mbar. Before the deposition, the chamber was evacuated down to $6\times10^{-6}$\,mbar. The elemental composition in the Co nanostripes is 75\,\%\,at. Co, 13\,\%\,at. O, and 12\,\%\,at. C as residues from the precursor, as inferred from energy-dispersive X-ray spectroscopy on thicker replica of the fabricated structures. Further details on the structural and magnetic properties of Co structures fabricated by FEBID can be found elsewhere \cite{Beg15nan}. To break the symmetry in the vortex dynamics under current reversal, the slopes of the nanostripes were designed asymmetric. In the patterning process, this was achieved by defining each nanostripe as a 5-step ``stair'' with a step-wise increasing number of FEBID passes assigned to their ``steps''. Due to blurring effects, smooth slopes of the nanostripes resulted instead of the ``stairs'', as inferred by atomic force microscopy (AFM), refer to Fig. 1(b) in the manuscript.

\textbf{Cryogenic spectroscopy}. The nanopatterned Nb CPWs have a superconducting transition temperature $T_c=8.05$\,K, as defined by the 90\% resistance criterion, and a room-to-10\,K resistance ratio of 4.4. The difference in $T_c$ of different CPWs does not exceed $0.1$\,K. The upper critical field for all samples at zero temperature $H_{c2}(0)$ is about $1.4$\,T, as deduced from fitting the dependence $H_{c2}(T)$ to the phenomenological law $H_{c2}(T) = H_{c2}(0)[1-(T/T_c)^2]$. This yields a Ginzburg-Landau coherence length $\xi(0) = \Phi_0^{1/2} [2 \pi H_{c2}]^{-1/2}\approx 15$\,nm, where $\Phi_0 = 2.07\times10^{-15}$\,Tm$^2$ is the magnetic flux quantum. Combined broadband microwave and dc voltage measurements were done in a $^4$He cryostat at the temperature $0.8T_c = 6.44$\,K with magnetic field $H$ directed perpendicular to the film surface. The samples were field cooled before each measurement. Though due to the complex cross-sectional profile of the Co stripes their magnetic state may differ from the single domain state (with magnetization directed along the stripe axis), we observed no effect on the pinning potential after the application of out-of-plane and in-plane fields of $2$\,T.

The samples were mounted in a copper housing in which pins of micro-SMP connectors were spring-loaded against the pre-formed 300\,nm-thick gold contact pads sputtered through a shadow mask onto the film surface after the nanopatterning step. The microwave signal was fed to the sample through coaxial cables from a vector network analyzer (VNA, Agilent E5071C). The microwave and dc signals were superimposed and uncoupled by using two bias-tees mounted at the VNA ports. The measured quantity is the forward transmission coefficient $S_{21}$, defined as a ratio (expressed in dB) of the microwave power measured at port 2 to the power transmitted at port 1. While $S_{21}$ is a complex quantity, here we use the notation $S_{21}$ for referring to its absolute value. The depinning frequency $f_\mathrm{d}$ can be determined from the complex transmission coefficient at the point where the phase difference between the viscous and pinning forces amounts to $\pi/2$. This definition is equivalent to the definition of $f_\mathrm{d}$ as the frequency of the maximum derivative of $IL$ with respect to $f$.\\[10mm]

\begin{acknowledgments}
OD acknowledges the German Research Foundation (DFG) for support through Grant No 374052683 (DO1511/3-1).
Support through the Frankfurt Center of Electron Microscopy (FCEM) is gratefully acknowledged.
This work was supported by the European Cooperation in Science and Technology via COST Action CA16218 (NANOCOHYBRI).
\end{acknowledgments}


%

\end{document}